\documentclass{aastex631} %[linenumbers]
\usepackage{booktabs}

\begin{document}

\title{Constraining the Origin of FRB 20121102A's Persistent Radio Source with Long-Term Radio Observations}

\author[0000-0002-3615-3514]{Mohit~Bhardwaj}
\affiliation{McWilliams Center for Cosmology, Department of Physics, Carnegie Mellon University, Pittsburgh, PA 15213, USA}
\correspondingauthor{Mohit Bhardwaj}
\email{mohitb@andrew.cmu.edu}

\author[0000-0003-0477-7645]{Arvind~Balasubramanian}
  \affiliation{Department of Astronomy and Astrophysics, Tata Institute of Fundamental Research, Mumbai, 400005, India}

\author[0000-0003-4382-4467]{Yasha Kaushal}
\affiliation{Department of Physics and Astronomy and PITT PACC,
University of Pittsburgh, 3941 O'Hara St, Pittsburgh, PA 15213, USA}

\author[0000-0003-2548-2926]{Shriharsh~P.~Tendulkar}
  \affiliation{Department of Astronomy and Astrophysics, Tata Institute of Fundamental Research, Mumbai, 400005, India}
  \affiliation{National Centre for Radio Astrophysics, Post Bag 3, Ganeshkhind, Pune, 411007, India}
  \affiliation{CIFAR Azrieli Global Scholars Program, MaRS Centre, West Tower, 661 University Ave, Suite 505, Toronto, ON, M5G 1M1 Canada}

\begin{abstract}
The persistent radio source (PRS) associated with FRB 20121102A, the first precisely localized repeating fast radio burst (FRB), provides key constraints on both its local environment and the nature of the underlying FRB engine. We present a seven-year (2016$–$2023) temporal analysis of the PRS, combining new uGMRT observations with archival data across L-band frequencies. We find no statistically significant long-term trend in its L-band flux density. The observed variability is consistent with refractive interstellar scintillation, and the data do not require the PRS to be a source exhibiting strong intrinsic variability. This stability challenges models predicting rapid fading from evolving magnetized outflows, such as those powered by young magnetars or relativistic shocks. Our low-frequency observations show no evidence for spectral evolution between 1.4 GHz and 745 MHz, with a measured spectral index of $\alpha = -0.15 \pm 0.08$, in agreement with values reported from earlier observations in 2016–2017. The PRS remains compact, exhibits a flat radio spectrum, and—if powered by an intermediate-mass black hole accreting at a low Eddington ratio—its radio and X-ray properties are broadly consistent with the fundamental plane of radio-loud AGNs. While not conclusive, this scenario represents a viable alternative to magnetar wind nebula models and warrants further investigation. Furthermore, we find no statistically significant correlation between FRB burst activity and the luminosity of associated PRSs among repeating sources. This apparent decoupling challenges simple progenitor models that directly link bursts and persistent emission. Together, these results suggest that the FRB engine and PRS may in some systems originate from physically distinct sources, underscoring the need for flexible models to explain the diverse environments of repeating FRBs.
\end{abstract}


\section{Introduction}
\label{sec:introduction}

Fast radio bursts (FRBs) are enigmatic extragalactic radio transients, characterized by their millisecond durations, extremely high brightness temperatures ($\sim 10^{35}$ K), and typical energies exceeding $10^{35}$\,erg \citep[e.g.,][]{2007Sci...318..777L,2022A&ARv..30....2P}. Despite a rapidly growing catalog of detected FRBs \citep{catalog12021ApJS,2023Univ....9..330X}, their precise physical origins remain a significant astrophysical enigma. While most FRBs are observed as singular events, a crucial subset has been found to repeat \citep[e.g.,][]{2023ApJ...947...83C}. These repeating FRBs offer unique opportunities for precise localization \citep[e.g.,][]{Marcote2017ApJ,Marcote_2020_Natur,2021ApJ...919L..24B,2021ApJ...910L..18B,Kirsten_2022_Natur,Nimmo_2022_ApJL,2023ApJ...950..134M,hewitt_2024_mnras_dec,2025arXiv250611915B}, multi-wavelength follow-up, and detailed studies of their host environments \citep[e.g.,][]{2022AJ....163...69B,2023ApJ...954...80G,2024ApJ...971L..51B,2024Natur.635...61S}, progenitor systems \citep[e.g.,][]{2014MNRAS.442L...9L,2017MNRAS.468.2726K,2018ApJ...868L...4M,2019ApJ...886..110M,2020ApJ...893L..39L,2020MNRAS.496.3390B,2020ApJ...892..135W,2021ApJ...917...13S,2021ApJ...922...98D,2024ApJ...977..122B,2025arXiv250621753B}, and potential persistent counterparts \citep[for detailed discussion, see][]{2021Univ....7...76N,2024ARNPS..74...89Z}.

FRB~20121102A, the first repeating fast radio burst discovered with the Arecibo telescope \citep{2016Natur.531..202S}, quickly became the most extensively studied, representing a significant breakthrough in our understanding of FRBs. Its precise localization to a low-metallicity, star-forming dwarf galaxy at a redshift of $z = 0.19273$ provided the first direct association between an FRB and its cosmic environment \citep{Tendulkar2017ApJ,2017ApJ...843L...8B}. Notably, the burst source was found to be co-spatial with a compact, luminous persistent radio source (hereafter PRS1). VLBI observations constrain PRS1 to be unresolved on milliarcsecond scales, implying a projected size of $\lesssim 0.7$\,pc \citep{Marcote2017ApJ}. Further constraints based on radio scintillation indicate a lower limit of $\gtrsim 0.03$\,pc and a most probable size of $\sim 0.2$\,pc \citep{Chen2023ApJ}. These size estimates yield a brightness temperature exceeding $10^7$\,K, confirming the non-thermal nature of the emission and ruling out standard thermal mechanisms (e.g., H\textsc{ii} regions or typical supernova remnants; \citealt{1992ARA&A..30..575C}). With a 1.4 GHz spectral luminosity of $L_{\nu} \approx 2 \times 10^{29}$\,erg\,s$^{-1}$\,Hz$^{-1}$, PRS1 far exceeds expectations from star formation in the host galaxy \citep{2017Natur.541...58C,Marcote2017ApJ}. Furthermore, deep optical and X-ray observations have not revealed any bright counterparts at those wavelengths \citep[e.g.,][]{Scholz2017ApJ,2017MNRAS.472.2800H,2018MNRAS.481.2479M}.

Regarding its radio spectrum, early observations revealed that PRS1 has a nearly flat spectral index between 400\,MHz and 6\,GHz, with a typical flux density around $200\,\mu$Jy across this band \citep{2017Natur.541...58C, Marcote2017ApJ, 2021AA...655A.102R, 2022MNRAS.511.6033P}. At higher frequencies, VLA observations show a decline in flux density \citep{2017Natur.541...58C}. This steepening initially prompted suggestions of a spectral cooling break near 10\,GHz \citep[e.g.,][]{2017Natur.541...58C}. However, later analysis using multi-band VLA observations found no compelling evidence for such a break when accounting for flux density variability induced by Galactic scintillation \citep{Chen2023ApJ}. At lower frequencies, particularly around 3\,GHz, PRS1 exhibits flux density modulations at the level of $\sim$25--30\% \citep{2017Natur.541...58C}. While such variability was initially attributed solely to refractive interstellar scintillation (RISS) in the Milky Way's interstellar medium, a subsequent analysis of multi-epoch data, primarily at 1.5 GHz, by \citet{2024ApJ...976..165Y} argued that the observed modulation index exceeded predictions from standard scintillation models and suggested the presence of intrinsic variability or the influence of a more complex scattering environment. 

More broadly, the association of compact, luminous persistent radio sources with repeating FRBs has since been established for a few other sources, including FRB~20190520B \citep{2022Natur.606..873N}, FRB~20201124A \citep{2024Natur.632.1014B}, and FRB 20240114A \citep{2024arXiv241213121B,2025AA...695L..12B}. These cases suggest a possible physical link between the repetition of the FRB and the presence of a PRS. However, for PRS1 itself, several independent studies have searched for correlations between its FRB activity and persistent radio flux density, and consistently found none \citep{2017Natur.541...58C,Marcote2017ApJ,2022MNRAS.511.6033P,Chen2023ApJ,2024ApJ...976..165Y}. The lack of any contemporaneous changes in PRS1 luminosity despite a clear, roughly monotonic decline in both the rotation measure (RM) and dispersion measure (DM) of the FRB source \citep{2021ApJ...908L..10H,2022MNRAS.511.6033P,2025ATel17156....1Z} further supports the interpretation that PRS1 may not be directly powered by individual burst activity. Instead, it could trace a separate, long-lived physical component unrelated to the FRB activity. Additionally, the limited number of such associations, alongside their absence in the majority of localized FRBs, indicates that PRSs may represent a distinct class of repeating FRB environments or a transient evolutionary phase, rather than a universal characteristic.

Reflecting its unusual luminosity and compactness, and considering the previously suggested but still debated aspects of its long-term temporal evolution, a wide range of theoretical models have been proposed to explain the origin of PRS1. The leading interpretation invokes a young millisecond magnetar powering a magnetized wind nebula (MWN), inflated by its relativistic outflows within a dense ionized medium \citep{2017ApJ...841...14M,2018MNRAS.481.2407M}. In these models, the synchrotron-emitting nebula can account for the observed radio luminosity and the extreme RM, with a gradual secular fading and spectral evolution expected over decade-long timescales. A variant of this scenario involves a magnetar embedded in a dense pre-explosion environment or pulsar wind bubble without surrounding supernova ejecta \citep{2017ApJ...838L...7D,2020ApJ...896...71L,2019ApJ...885..149Y}. Another proposed class of models suggests that the PRS arises from an off-axis afterglow of a long-duration gamma-ray burst or a superluminous supernova (LGRB/SLSN), where the jet does not point toward Earth but its late-time isotropic emission remains detectable in radio \citep{2018MNRAS.481.2407M,2019ApJ...876L..14M,2019ApJ...876L..10E}. Alternatively, the PRS has been linked to a low-luminosity AGN, potentially a radiatively inefficient accretion flow onto a massive black hole \citep{Marcote2017ApJ,2017A&A...602A..64V,2020ARA&A..58..257G,2020ApJ...895...98E}. A different class of models invokes hyperaccreting compact binaries (e.g., ultra-luminous X-ray binaries or microquasars) which can drive powerful synchrotron radio nebulae in a dense circumstellar environment \citep{Sridhar2022ApJ}. Other exotic proposals include cosmic comb models \citep{2017ApJ...836L..32Z} and generic synchrotron-heated nebulae \citep{2024arXiv240507446C}, though these are less directly constrained by current data. Continued monitoring of the radio spectrum, variability, and polarization, along with multi-wavelength constraints, will be crucial to discriminating among these models.

While some previous long-term monitoring campaigns have suggested evidence for intrinsic flux density variations and a gradual decline of PRS1's luminosity up to certain epochs \citep[e.g.,][]{2023MNRAS.525.3626R}, the overall long-term evolution of PRS1 remains an active and critical area of investigation. Unraveling this behavior requires long-term multi-frequency monitoring to decisively distinguish between competing astrophysical scenarios. In this work, we present new long-term observations of PRS1 using the upgraded Giant Metrewave Radio Telescope (uGMRT) at central frequencies of 400\,MHz, 650\,MHz, and 1.4\,GHz. Our new data extend the multi-frequency baseline for PRS1 through 2023, enabling a comprehensive investigation into its long-term spectral and temporal evolution. These unique low-frequency constraints are crucial for distinguishing between evolving astrophysical scenarios for the PRS and for probing the frequency dependence of any intrinsic variability or environmental effects. Our observations provide updated constraints on PRS1's long-term radio flux density, revealing its behavior across recent epochs and allowing for a thorough reassessment of its overall evolution. The remainder of this paper is organized as follows: \S\ref{sec:obs_data_analysis} describes the observational setup, data reduction, measured flux densities, and spectral properties. In \S\ref{sec:discussion}, we discuss the implications of our results in the context of theoretical models. Finally, \S\ref{sec:conclusion} summarizes our conclusions.

\section{Observations and Data Analysis}
\label{sec:obs_data_analysis}

This section details the radio observations of PRS1, including the methodologies for data acquisition and reduction, and the subsequent analysis of its long-term temporal evolution and spectral characteristics.

\subsection{GMRT Flux Density Measurements}
\label{subsec:gmrt_flux_measurements}

We conducted new uGMRT observations of PRS1 across three epochs: 01 August 2023, 08 August 2023, and 16 August 2023. These observations were part of project 44\_032 (PI: Bhardwaj) and utilized the wideband backend receiver system in three frequency bands: Band 3 (centered at 400\,MHz, 200\,MHz bandwidth), Band 4 (centered at 745\,MHz, 400\,MHz bandwidth), and Band 5 (centered at 1264\,MHz, 400\,MHz bandwidth). Additionally, we incorporated two earlier uGMRT Band 5 observations (29 November 2022 and 21 March 2023) from program 43\_054 (PI: Yi Feng).

Data were downloaded in \texttt{FITS} format and processed using \texttt{CASA} \citep{2022PASP..134k4501C}. Calibration and imaging were performed using \texttt{CASA-CAPTURE}, an automated continuum imaging pipeline for CASA \citep{2021ExA....51...95K}, with iterative self-calibration (typically eight rounds) applied to enhance image fidelity. For project 44\_032, 3C147 was used as the primary flux and bandpass calibrator, and 0431+206 as the phase calibrator. For project 43\_054, 3C147 and 0555+398 served as the bandpass and phase calibrators, respectively. The absolute flux scale was referenced to the Perley-Butler 2017 model of 3C147 \citep{2017ApJS..230....7P}. Flux density measurements of PRS1 were obtained using the \texttt{CASA} task \texttt{imfit}, applied to a circular region centered on the source. We added a 10\% systematic uncertainty in quadrature to the statistical \texttt{imfit} errors to account for known calibration uncertainties in GMRT data, particularly at low frequencies \citep[e.g.,][]{2021AA...655A.102R}. As a consistency check, the measured flux densities of the calibrator 3C147 were compared to the expected values and are reported in Table~\ref{tab:gmrt_obs_results}.

In Band 5, PRS1 was clearly detected at all epochs with measured flux densities ranging from 228--297\,$\mu$Jy, consistent with long-term flux stability (see \S\ref{subsec:long_term_varability}). In Band 4, we also detected PRS1 at all three epochs, with flux densities ranging from 230--295\,$\mu$Jy. These values are consistent with archival detections at similar frequencies, including $231 \pm 22\,\mu$Jy at 668\,MHz \citep{2020MNRAS.498.3863M} and $276.5 \pm 69\,\mu$Jy at 610\,MHz \citep{2021AA...655A.102R}, suggesting spectral and temporal stability in the 550–800\,MHz range. We quantify variability and compute an average flux density across this band in \S\ref{subsec:long_term_varability}.

In Band 3, PRS1 was not detected at any epoch. We derive $3\sigma$ upper limits of 275–295\,$\mu$Jy at 400\,MHz, which are consistent with but less constraining than earlier detections at similar frequencies: $203.5 \pm 33.6\,\mu$Jy at 400\,MHz \citep{2021AA...655A.102R} and $215 \pm 37\,\mu$Jy at 433\,MHz \citep{2020MNRAS.498.3863M}. These upper limits do not provide strong evidence for spectral curvature or absorption within the observed bandpass, but are consistent with PRS1's flat spectrum as noted by \citet{2021AA...655A.102R}.

Finally, all flux density measurements and observing parameters are listed in Table~\ref{tab:gmrt_obs_results}.

\begin{table}[h]
\centering
\footnotesize
\rmfamily
\begin{tabular}{c|c|c|c|c|c}
\hline
\bf Epoch & \bf Central Frequency & \bf Flux Density PRS1 & \bf Bandpass Calibrator & \bf Obs. Time & Proposal ID \\
\bf (YYYY-MM-DD) & \bf (MHz) & \bf ($\mu$Jy) & \bf (Jy) & \bf (Hrs) & \\
\hline\hline
16 August 2023 & 400 & $<$295  & -- & 1.1 & 44\_032 \\ 
01 August 2023 & 400 & $<$275 & -- & 1.1 & 44\_032 \\ 
08 August 2023 & 400 & $<$275 & -- & 1.1 & 44\_032 \\ 
16 August 2023 & 745 & $295 \pm 42$ & $34.33 \pm 0.04$ & 0.5 & 44\_032 \\ 
08 August 2023 & 745 & $258 \pm 44$ & $34.30 \pm 0.03$ & 0.5 & 44\_032 \\ 
01 August 2023 & 745 & $230 \pm 46$ & $34.69 \pm 0.07$ & 1.1 & 44\_032 \\ 
01 August 2023 & 1264 & $256 \pm 29$ & $23.87 \pm 0.01$ & 1.1 & 44\_032 \\ 
16 August 2023 & 1264 & $229 \pm 27$ & $20.86 \pm 0.06$ & 0.5 & 44\_032 \\ 
08 August 2023 & 1264 & $228 \pm 28$ & $21.00 \pm 0.06$ & 0.5 & 44\_032 \\
\hline
29 November 2022 & 1265 & $219 \pm 30$ & $23.76 \pm 0.03$ & 0.4 & 43\_054 \\
21 March 2023 & 1260 & $297 \pm 42$ & $23.55 \pm 0.04$ & 0.4 & 43\_054 \\
\hline
\end{tabular}
\caption{uGMRT observations of PRS1. The first three entries are non-detections in Band~3 and quoted as $3\sigma$ upper limits.}
\label{tab:gmrt_obs_results}
\end{table}

\begin{table}[h!]
\footnotesize
\rmfamily
\begin{center}
\begin{tabular}{c|c|c|c|c|c}
\hline
\bf Date & \bf MJD & \bf Telescope & \bf Central Frequency & \bf Flux Density & \bf Reference \\
& & & \bf (MHz) & \bf ($\mu$Jy) & \\
\hline
\hline
02 February 2016 & 57420 & EVN & 1670 & 200 $\pm$ 20 & \cite{2017Natur.541...58C} \\
10 February 2016 & 57428 & EVN & 1690 & 200 $\pm$ 20 & \cite{Marcote2017ApJ} \\
11 February 2016 & 57429 & EVN & 1690 & 175 $\pm$ 14 & \cite{Marcote2017ApJ} \\
24 May 2016 & 57532 & EVN & 1690 & 220 $\pm$ 40 & \cite{Marcote2017ApJ} \\
25 May 2016 & 57533 & EVN & 1690 & 180 $\pm$ 40 & \cite{Marcote2017ApJ} \\
07 September 2016 & 57638 & VLA & 1630 & 250 $\pm$ 39 & \cite{2017Natur.541...58C} \\
09 September 2016 & 57640 & VLBA & 1550 & 218 $\pm$ 38 & \cite{2017Natur.541...58C} \\
20 September 2016 & 57651 & EVN & 1690 & 168 $\pm$ 11 & \cite{Marcote2017ApJ} \\
23 February 2017 & 57807 & EVN & 1700 & 239 $\pm$ 62 & \cite{2022MNRAS.511.6033P} \\
13 May 2017 & 57886 & EVN & 1700 & 278 $\pm$ 54 & \cite{2022MNRAS.511.6033P} \\
20 May 2017 & 57893 & GMRT & 1390 & 148.5 $\pm$ 60 & \cite{2021AA...655A.102R} \\
03 November 2017 & 58060 & EVN & 1700 & 232 $\pm$ 32 & \cite{2022MNRAS.511.6033P} \\
10 December 2017 & 58097 & GMRT & 1260 & 241.5 $\pm$ 11.1 & \cite{2021AA...655A.102R} \\
06 September 2019 & 58732 & MeerKAT & 1280 & 260 $\pm$ 26 & \cite{2023MNRAS.525.3626R} \\
10 September 2019 & 58736 & MeerKAT & 1280 & 269 $\pm$ 27 & \cite{2023MNRAS.525.3626R} \\
06 October 2019 & 58762 & MeerKAT & 1280 & 287 $\pm$ 27 & \cite{2023MNRAS.525.3626R} \\
08 October 2019 & 58764 & MeerKAT & 1280 & 270 $\pm$ 28 & \cite{2023MNRAS.525.3626R} \\
26 September 2022 & 59848 & MeerKAT & 1280 & 189 $\pm$ 18 & \cite{2023MNRAS.525.3626R} \\
29 November 2022 & 59912 & GMRT & 1265 & 219 $\pm$ 30 & this work \\
21 March 2023 & 60024 & GMRT & 1260 & 297 $\pm$ 42 & this work \\
28 May 2023 & 60092 & VLA & 1500 & 264.7$\pm$52.2 & \cite{2024ApJ...976..165Y} \\
29 May 2023 & 60093 & VLA & 1500 & 361.6$\pm$56.4 & \cite{2024ApJ...976..165Y} \\
05 June 2023 & 60100 & VLA & 1500 & 224.3$\pm$41.6 & \cite{2024ApJ...976..165Y} \\
13 June 2023 & 60108 & VLA & 1500 & 234.6$\pm$47.8 & \cite{2024ApJ...976..165Y} \\
18 June 2023 & 60113 & VLA & 1500 & 325.7$\pm$48.0 & \cite{2024ApJ...976..165Y} \\
18 June 2023 & 60113 & VLA & 1500 & 300.4$\pm$57.3 & \cite{2024ApJ...976..165Y} \\
19 June 2023 & 60114 & VLA & 1500 & 269.2$\pm$59.5 & \cite{2024ApJ...976..165Y} \\
22 June 2023 & 60117 & VLA & 1500 & 206.6$\pm$40.5 & \cite{2024ApJ...976..165Y} \\
26 June 2023 & 60121 & VLA & 1500 & 222.5$\pm$28.7 & \cite{2024ApJ...976..165Y} \\
27 June 2023 & 60122 & VLA & 1500 & 110.3$\pm$22.7 & \cite{2024ApJ...976..165Y} \\
01 August 2023 & 60157 & GMRT & 1264 & 256 $\pm$ 29 & this work \\
08 August 2023 & 60164 & GMRT & 1264 & 228 $\pm$ 28 & this work \\
16 August 2023 & 60172 & GMRT & 1264 & 229 $\pm$ 27 & this work \\
\hline
\end{tabular}
\end{center}
\caption{Summary of L-band PRS1 Flux Density Measurements. This table includes observations from various telescopes and frequencies, covering the period from 2016 to 2023. Our new uGMRT measurements are indicated as "this work." All measurements were made at L-band frequencies, ranging from 1260\,MHz to 1700\,MHz. Due to the known flat spectrum of the PRS and the negligible change in its modulation index across these frequencies, all values were included in our long-term variability study of the PRS radio light curve, as shown in Figure \ref{fig:lband_lightcurve}. Flux density values and their associated errors are taken from the cited references.}
\label{tab:flux_density_measurements_lband}
\end{table}

\subsection{Long-term Radio Temporal Variability of PRS1}
\label{subsec:long_term_varability}

To investigate the long-term temporal variability of PRS1, we combined our new uGMRT observations (Table~\ref{tab:gmrt_obs_results}) with extensive archival data from various radio telescopes, including the VLA, EVN, MeerKAT, and GMRT, spanning from 2016 to 2023. We focus our variability analysis on L-band frequencies (1260–1700 MHz), as this range provides the most extensive coverage across instruments, with a total of 33 flux density measurements available. These combined L-band measurements are summarized in Table~\ref{tab:flux_density_measurements_lband} and visualized in Figure~\ref{fig:lband_lightcurve}.  Given the flat radio spectrum of PRS1 across this frequency range \citep{2017Natur.541...58C}, we expect negligible intrinsic spectral evolution, allowing us to treat these measurements as representative of the same emission component.

To quantify the long-term variability, we first utilized the modulation index ($m$) and the weighted reduced chi-square statistic ($\eta$), which characterize the amplitude and significance of flux variations, respectively \citep[e.g.,][]{Sarbadhicary2021}:
\begin{equation} \label{eq:mod_index}
    m = \frac{1}{\overline{F}} \sqrt{\frac{N}{N-1} \left(\overline{F^2} - \overline{F}^2\right)},
\end{equation}
\begin{equation}
 \text{and}~\eta = \frac{1}{N-1} \sum_{i=1}^{N} \frac{(F_i - \xi_F)^2}{\sigma_i^2},
\label{eq:chi_square}
\end{equation}

where $\overline{F}$ is the unweighted mean flux density, $N$ is the number of measurements, $F_i$ is the $i$-th flux density measurement with uncertainty $\sigma_i$, and $\xi_F$ is the weighted mean.

We find a modulation index of $m = 0.22$ and a reduced chi-square value of $\eta = 3.18$ for the 33 flux density measurements, corresponding to a $p$-value of $3.5 \times 10^{-9}$ for the null hypothesis of constant flux. This indicates that the observed variations are statistically significant and inconsistent with being solely due to measurement noise.

We then evaluated whether the observed variability could be attributed to the RISS. Scintillation, caused by the propagation of radio waves through a turbulent ionized ISM \citep{Rickett1990}, can induce flux variability. We assessed the role of Galactic scintillation by applying the framework described by \cite{1998MNRAS.294..307W}. Using the NE2001 Galactic electron density model \citep{Cordes2002}, we estimated the transition frequency ($\nu_t$) between weak and strong scattering for the line of sight to PRS1 to be $\nu_t \approx 38$\,GHz. In the strong scattering regime ($\nu \ll \nu_t$), the modulation index for RISS is expected to scale with frequency as
\begin{equation}
m_{\mathrm{RISS}} \approx \left( \frac{\nu}{\nu_t} \right)^{17/30},
\end{equation}
which yields $m_{\mathrm{RISS}} \approx 0.15$ at 1.4\,GHz. The corresponding refractive scintillation timescale is given by \citep{1998MNRAS.294..307W}
\begin{equation}
\tau_{\mathrm{RISS}} \approx 2 \left( \frac{\nu_t}{\nu} \right)^{11/5} \,\mathrm{hr},
\end{equation}
yielding $\tau_{\mathrm{RISS}} \approx  117$\,days. This timescale represents the typical interval over which refractive scintillation induces significant flux variations; measurements obtained within shorter intervals are likely to be temporally correlated and not statistically independent.

To account for such temporal correlations, we implemented a Monte Carlo framework that resampled the full light curve under the assumption that only flux measurements separated by more than 117 days represent independent realizations of the intrinsic flux. In each of the 10,000 realizations, we selected a subset of data points meeting the temporal spacing criterion, requiring at least five measurements to ensure a reliable estimate of variability. This threshold was chosen to mitigate temporal correlations introduced by refractive scintillation while retaining enough data points for statistically meaningful analysis. For each realization, we computed the modulation index while incorporating flux density uncertainties. The resulting distribution yields a mean simulated modulation index of $\langle m \rangle = 0.210 \pm 0.085$. This simulated value ($\langle m \rangle = 0.210 \pm 0.085$) is consistent within $1\sigma$ of the observed modulation index ($m = 0.22$). This suggests that the observed variability can be fully explained by RISS and measurement uncertainties, implying that PRS1 does not exhibit significant intrinsic flux variations over the 7-year monitoring baseline.

To further assess flux variability at low frequencies, we combined our new Band 4 detections with prior measurements from \citet{2020MNRAS.498.3863M,2021AA...655A.102R} at 610–668 MHz. Using five measurements spanning 550–800 MHz, we find a modulation index of $m = 0.11 \pm 0.08$, consistent with being entirely due to measurement noise and expected RISS variations. This modulation level is in agreement with RISS predictions at 745 MHz using the \citet{1998MNRAS.294..307W} formalism. The inverse-variance weighted mean flux density across this band is $246 \pm 16$ $\mu$Jy, which we adopt for subsequent spectral index calculations.

Finally, we note that diffractive interstellar scintillation (DISS) is not expected to affect the flux densities reported here. The estimated scintillation bandwidth at 1\,GHz using NE2001 model is $\Delta\nu_{\mathrm{DISS}} \sim 8.4$\,kHz, much smaller than the typical observing bandwidths ($\gtrsim 10$\,MHz) of our measurements, which average over narrowband fluctuations.

\subsection{Assessment of Long-Term Trends}
\label{subsec:overall_trend}

Following our analysis of long-timescale variability, we examined whether the L-band flux density of PRS1 shows any long-term secular trend between 2016 and 2023. To ensure robustness, we employed three complementary statistical tests: (1) linear regression to identify linear trends, (2) the non-parametric Mann–Kendall test to detect monotonic trends, and (3) the Theil–Sen estimator \citep{theil1950a,sen1968}, which provides a slope estimate that is robust to outliers. The linear regression, implemented using \texttt{scipy} \citep{2020NatMe..17..261V}, yields a slope of $5.28 \pm 2.80~\mu$Jy\,yr$^{-1}$ with a $p$-value of 0.065, indicating that the trend is not statistically significant. The Mann–Kendall test, performed with \texttt{pymannkendall} \citep{Hussain2019}, returns a $p$-value of 0.077 and a Kendall’s Tau statistic of 0.22. The Theil–Sen estimator similarly yields a slope consistent with zero within uncertainties. All three tests fail to detect a statistically significant trend at the 5\% level. Over the 7-year monitoring period, the inverse-variance weighted mean flux density is $213 \pm 4~\mu$Jy. Taken together, these results show no compelling evidence for long-term brightening or fading of PRS1, reinforcing the interpretation that its L-band emission has remained remarkably stable over time.

\section{Discussion}
\label{sec:discussion}

Our new uGMRT observations presented in \S\ref{subsec:gmrt_flux_measurements}, combined with extensive archival data spanning from 2016 to 2023, provide the most comprehensive long-term view to date of the radio temporal evolution of PRS1. We find no statistically significant evidence for a long-term trend in its L-band flux density (\S\ref{subsec:overall_trend}), with an inverse-variance weighted mean flux density of $213 \pm 4~\mu$Jy.

Although the source is broadly stable over multi-year timescales, we detect statistically significant flux variability at 1.4\,GHz (\S\ref{subsec:long_term_varability}). A detailed Monte Carlo analysis accounting for the RISS and temporal correlations demonstrates that the observed modulation index ($m = 0.22$) is fully consistent with RISS-induced variability, particularly when the effects of sparse time sampling are taken into account. This supports the conclusion that PRS1 does not exhibit strong intrinsic variability.

These findings are further supported by our low-frequency observations. At 745\,MHz, we detect modest variations in flux density that are statistically consistent with measurement noise and expected RISS contributions. The modulation index across five measurements in the 550--800\,MHz range is $m = 0.11 \pm 0.08$, in good agreement with the predicted RISS amplitude at this frequency. The inverse-variance weighted mean flux density at 745\,MHz is $246 \pm 16~\mu$Jy. When compared with the L-band mean flux density of $213 \pm 4~\mu$Jy, this yields a spectral index of $\alpha = -0.15 \pm 0.08$ (where $F_\nu \propto \nu^\alpha$), consistent with a flat or mildly declining spectrum, as noted in previous studies.

In addition, we obtained new upper limits at 400\,MHz from three epochs in 2023, all of which lie above previous detections at similar frequencies \citep{2020MNRAS.498.3863M, 2021AA...655A.102R}. These non-detections are consistent with past flux measurements and do not show evidence for spectral turnover or curvature at the lowest observed frequencies. However, deeper observations at 400\,MHz and below may be required to test for potential free-free absorption or other low-frequency spectral suppression mechanisms.

In the following subsections, we examine the implications of these results for various physical models proposed for PRS1 and its possible connection to the FRB progenitor.

\begin{figure}[h!]
\begin{center}
\includegraphics[width=0.9\textwidth]{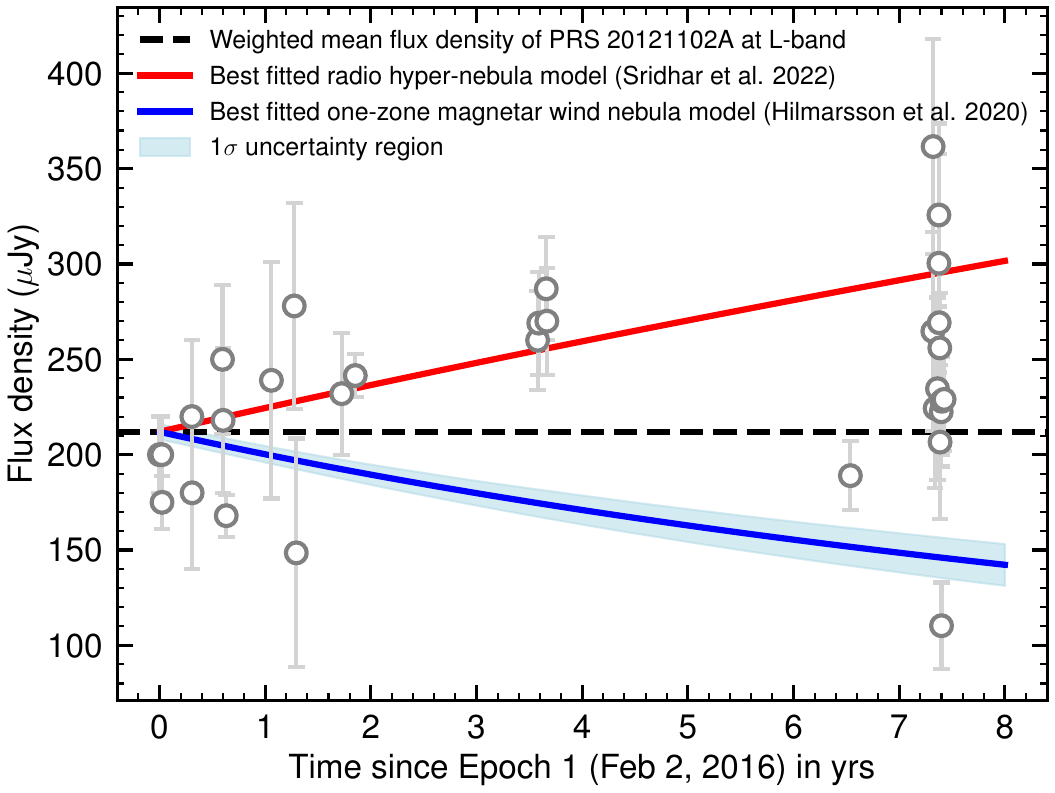}
\end{center}
\caption{The radio light curve of the persistent radio source associated with FRB~20121102A at L-band (data detailed in Table \ref{tab:flux_density_measurements_lband}). The dotted horizontal line represents the inverse-variance weighted mean flux density of the PRS, which is 213 $\mu$Jy. Overplotted are the best-fitted model predictions for the temporal evolution from the Magnetar Wind Nebula model (dashed red line) and Hypernebula model (solid blue line), normalized to the observed weighted mean flux at their respective inferred ages (as discussed in \S\ref{subsubsec:mwn_model} and \S\ref{subsubsec:hypernebula_model}). This figure illustrates the tension between the observed long-term stability and the predicted temporal evolution of these models.}
\label{fig:lband_lightcurve}
\end{figure}

\subsection{Evaluating FRB-PRS Models}
\label{subsec:model_evaluation}

The absence of a statistically significant long-term trend in flux density, combined with the modest level of variability that can be fully accounted for by refractive interstellar scintillation, provides meaningful constraints on the nature of PRS1. Any viable model for the persistent emission associated with FRB~20121102A must therefore accommodate a stable radio luminosity over at least a seven-year baseline and explain the lack of pronounced intrinsic variability at gigahertz frequencies. In this section, we evaluate three prominent models based on their ability to reproduce PRS1's observed properties.

\subsubsection{Magnetar Wind Nebula Model}
\label{subsubsec:mwn_model}

The magnetar wind nebula model proposes that PRS1 arises from synchrotron emission produced by a nebula inflated by the relativistic wind of a young, flaring magnetar \citep{2017ApJ...841...14M}. In this scenario, the magnetar injects magnetic energy and relativistic particles into a confined ion--electron nebula, which drives its expansion and powers the observed radio luminosity. If the energy injection declines as a power law in time, $\dot{E}_B(t) \propto t^{-\alpha}$, the resulting synchrotron flux density evolves as
\begin{equation}
F_\nu(t) \propto t^{-p}, \quad \text{with} \quad p = \frac{\alpha^2 + 7\alpha - 2}{4},
\end{equation}
as derived by \citet{2018ApJ...868L...4M}.

The MWN scenario has gained support from its success in reproducing the observed evolution of the RM and DM of FRB~20121102A. In particular, \citet{2021ApJ...908L..10H} modeled these quantities using the MWN formalism and found that the best-fit solutions require a nebula age of $t_{\rm age} \approx 15$--17\,yr and energy injection index $\alpha \approx 1.1$--1.6. However, applying these same parameters to the flux evolution of PRS1 leads to a discrepancy. Over a 7-year monitoring baseline, this range of $\alpha$ predicts decay indices of $p \approx 1.5$--2.9, corresponding to declines in L-band flux density of approximately 35--60\%. As shown in Figure~\ref{fig:lband_lightcurve} and discussed in \S\ref{sec:obs_data_analysis}, the PRS1 flux remains statistically consistent with a constant value of $213 \pm 4\,\mu$Jy over this period. A linear regression analysis yields a best-fit slope of $+5.3 \pm 2.8~\mu$Jy\,yr$^{-1}$, and a suite of non-parametric trend tests also fail to identify any significant monotonic change (\S\ref{subsec:overall_trend}). We also account for the effects of the RISS, which introduces long-timescale correlated variability that can obscure secular flux trends. As discussed in \S\ref{subsec:long_term_varability}, we estimate a refractive modulation index of $m_{\rm RISS} \approx 0.15$ and a timescale of $\sim$117 days. Incorporating this into the flux trend analysis yields an additional uncertainty in the slope of $\sim$1.7\,\textmu Jy\,yr$^{-1}$, raising the total uncertainty to $\sim$3.3\,\textmu Jy\,yr$^{-1}$.

For a given $\alpha$, we can translate the predicted flux decay into a linear slope, $dF/dt = -pF_0/t$, and compare this to the observed value. For $\alpha = 1.1$ ($p \approx 1.5$), nebular ages younger than $\sim$57\,yr are excluded at the 2$\sigma$ level; for $\alpha = 1.6$ ($p \approx 2.9$), the required minimum age increases to $\sim$110\,yr. These values are significantly older than the $\sim$15--17\,yr age inferred from RM and DM evolution, highlighting a potential inconsistency with the best-fit one-zone MWN interpretation. We emphasize that the MWN model includes multiple tunable parameters—such as magnetic energy partition, density profile, and injection history \citep{2018ApJ...868L...4M}—that could, in principle, produce flatter light curves under different conditions. However, such adjustments must simultaneously explain the observed RM and DM evolution of FRB~20121102A, which places strong constraints on the allowable parameter space.

Independent constraints on the FRB emission region pose additional challenges for the MWN scenario. \citet{Snelders2023NatAs} recently reported bursts from FRB~20121102A with durations $\lesssim 15\,\mu$s and rise times as short as 1\,$\mu$s, implying emission regions smaller than 300 meters. These ``ultra-FRBs'' exhibit polarimetric properties consistent with longer-duration bursts \citep{Hessels2019}, suggesting a shared emission mechanism. The extreme temporal and spatial coherence favors a magnetospheric origin within the light cylinder, and disfavors models where bursts are produced at large radii (e.g., $\gtrsim 10^{13}$\,cm) through synchrotron maser shocks or magnetic reconnection in relativistic outflows. For an observed rise time of $\sim$1\,$\mu$s, the causally connected emission region must be smaller than $R \lesssim c \tau \sim 300$\,m, in stark contrast to the much larger emission radii typically invoked in far-field models. Such models generally require low-density environments for the bursts to escape, which appears inconsistent with the dense, magnetized medium inferred from the large and time-variable RM of FRB~20121102A. 

In summary, the lack of measurable flux evolution in the PRS, when combined with the compact burst emission and high RM variability, places strong constraints on the standard MWN interpretation. The combined data disfavor a young MWN as the sole origin of the PRS and suggest that either the source is substantially older than inferred from RM evolution, or that additional physical ingredients—such as spatial gradients, multiple zones, or intermittent energy injection—are needed to reconcile the full set of observables.

\subsubsection{Hypernebula Model}
\label{subsubsec:hypernebula_model}

Another model proposed to explain the origin of PRS1 is the ``hypernebula'' scenario, proposed by \citet{Sridhar2022ApJ}. In this framework, the PRS is powered by synchrotron emission from a nebula inflated by a relativistic outflow launched during a short-lived hyperaccretion phase involving a compact object embedded within a common-envelope binary. The resulting nebula expands and shocks the surrounding dense medium, producing long-lasting radio emission. Concurrently, FRBs are generated at large distances ($\gtrsim 10^{13}$\,cm) from the engine via coherent maser emission in relativistic internal shocks. 

In contrast to the MWN model, the hypernebula scenario allows for a modest increase in PRS luminosity during its early expansion. For FRB~20121102A, \citet{Sridhar2022ApJ} identified a best-fit solution with a nebular age of $t_{\rm age} \approx 10$\,yr and a flux evolution trend of $L_\nu \propto t^{0.6}$. We overplot this model in Figure~\ref{fig:lband_lightcurve}, normalized to match the observed inverse-variance weighted flux density of $213 \pm 4\,\mu$Jy at $t = 10$\,yr. Over the 7-year baseline of our monitoring campaign, the model predicts a flux increase of $\sim$90\,$\mu$Jy, or $\sim$42\%, reaching $\sim$303\,$\mu$Jy at $t = 17$\,yr. Even if we take the best-fit linear regression slope of $+5.3 \pm 2.8\,\mu$Jy\,yr$^{-1}$ (\S\ref{subsec:overall_trend}) at face value, the observed increase falls short of the predicted slope of $+11.6\,\mu$Jy\,yr$^{-1}$ from the hypernebula model. The difference corresponds to a $\sim$1.9$\sigma$ discrepancy. When the effects of RISS are included—which increase the effective measurement uncertainty by inducing long-timescale temporal correlations—the uncertainty in the slope rises to $\sim$3.3\,$\mu$Jy\,yr$^{-1}$, marginally reducing the statistical tension.

While the best-fit hypernebula model with $t_{\rm age} = 10$\,yr and $L_\nu \propto t^{0.6}$ predicts a modest but measurable rise in PRS flux, our observations show no statistically significant long-term increase. The resulting tension is mild, and when the effects of RISS are included, the model becomes only marginally inconsistent with the data. Notably, the hypernebula framework admits a broader parameter space, and modest adjustments to input assumptions could, in principle, reconcile the observed flux stability with this scenario. Thus, while our results mildly disfavor the specific best-fit hypernebula evolution, they do not rule out the model as a whole.

However, like the MWN model, the hypernebula scenario relies on FRB emission originating at large radii, whereas recent observations of ultra-fast burst durations from FRB~20121102A \citep{Snelders2023NatAs} favor emission regions located within the magnetosphere of a compact object. Producing such temporally coherent bursts in far-field maser shock models would require finely tuned conditions, particularly in the dense environments expected in the hypernebula scenario. Therefore, while the flux evolution of PRS1 does not strongly disfavor the hypernebula model, the burst properties themselves pose a significant challenge to its viability.

\subsubsection{Relativistic Shock Models: LGRB/SLSN Afterglow and Supernova Ejecta}
\label{subsubsec:relativistic_shock_models}

Another proposed origin for PRS1 is synchrotron emission from an external shock, powered either by a relativistic jet launched during a long gamma-ray burst or superluminous supernova event \citep[e.g.,][]{2016MNRAS.461.1498M,2019ApJ...876L..10E}, or by a slower non-relativistic shock propagating into the surrounding circumstellar medium (CSM) following a supernova explosion \citep{Chevalier1998ApJ}. In these models, the persistent radio emission is generated by forward-shock-accelerated electrons radiating in compressed magnetic fields.

However, several observations disfavor this class of scenarios in the context of FRB~20121102A:

First, the radio luminosity of PRS1 at 1.4\,GHz, $L_\nu \sim 2 \times 10^{29}$\,erg\,s$^{-1}$\,Hz$^{-1}$, significantly exceeds that of well-studied Type Ib/c SNe and GRB afterglows at comparable epochs \citep{1989ApJ...336..421W,2009AJ....138.1529U,2019A&A...623A.173V,2020MNRAS.498.3863M}. Even for off-axis jets or dense CSM, producing such high luminosity at $\gtrsim 10$\,yr post-explosion typically requires unusually energetic outflows or extreme environmental conditions that verge on being fine-tuned.

Second, the flux evolution presents a challenge. In standard afterglow models for adiabatic expansion into a uniform medium and observing frequencies $\nu > \nu_m$, the synchrotron flux scales approximately as $F_\nu \propto t^{-1.2}$ to $t^{-1.5}$ at late times \citep{2002ApJ...568..820G}. This predicts a $\gtrsim 3$-fold decline over 7 years, in stark contrast to the observed stability of PRS1. Such plateauing would require either the jet to remain narrowly collimated and misaligned for over a decade, or for the shock to be continuously refreshed---possibilities not obviously supported by accompanying signatures in the radio SED or variability.

Third, the free--free transparency of the medium provides an additional age constraint. The detection of bursts from FRB~20121102A down to $\sim 600$\,MHz \citep{2019ApJ...882L..18J} implies that the surrounding ionized ejecta is optically thin to low-frequency radio waves, requiring $t_{\rm age} \gtrsim 15$--20\,yr for typical ejecta densities and ionization profiles \citep{2018ApJ...868L...4M}. At such ages, typical shock-powered synchrotron emission would have faded significantly unless supplemented by ongoing energy injection.

Moreover, the RM of $\sim 10^5$\,rad\,m$^{-2}$ associated with the bursts is hard to reconcile with the more extended and lower-density environments of external shocks \citep{2020MNRAS.498.3863M}. In LGRB or SLSN remnants, the forward shock is not expected to generate such high RM values unless coincidentally embedded within a compact magnetized nebula, thus invoking a multi-component geometry where the FRB and PRS trace distinct physical regions---a contrived and less parsimonious arrangement.

Finally, the observed monotonic decline in both RM and DM of FRB~20121102A without a corresponding decline in PRS flux argues against a single-phase model (e.g., CSM interaction) being responsible for all three observables. A decoupling of the burst and persistent emission regions becomes necessary, undermining the coherence of the shock model.

Taken together, these arguments suggest that while external shock models remain plausible for other PRS candidates, especially those associated with transient events like PTF10hgi \citep{2019ApJ...876L..10E}, they are increasingly disfavored in the case of PRS1. The energetic demands, temporal evolution, and environmental constraints all point toward a more compact and persistent engine, such as a young magnetar wind nebula or a hypernebula, as the likely origin.

\subsubsection{Low-Luminosity Active Galactic Nucleus (AGN) Scenario}
\label{subsubsec:agn_scenario}

The AGN scenario posits that PRS1 originates from synchrotron emission produced by a compact active galactic nucleus powered by accretion onto an intermediate-mass black hole (IMBH), either through the accretion disk or at the base of a jet \citep{Marcote2017ApJ, 2017ApJ...836L..32Z, 2017A&A...602A..64V}. Our finding of long-term L-band flux stability, coupled with low-level variability on multi-year timescales, is broadly consistent with the behavior of compact, low-luminosity radio-loud AGNs. Although long-term monitoring of such sources in dwarf galaxies is limited, instances of multi-year stability or low-level variability have been reported among low-luminosity radio AGNs \citep[e.g.,][]{2000ApJ...542..186N, 2004NewAR..48.1157F, 2005ApJ...627..674A, 2011MNRAS.412.2641J, 2015A&A...576A..38B, 2023AARv..31....3B}.

PRS1 exhibits several properties that are qualitatively consistent with AGN cores: it is compact ($< 0.35$\,pc), has a flat or slightly inverted spectral index ($\alpha \gtrsim -0.2$), and possesses a radio luminosity of $L_\nu \sim 10^{29}$\,erg\,s$^{-1}$\,Hz$^{-1}$ at 1.4\,GHz. These characteristics are reminiscent of face-on, core-dominated jet-mode AGNs \citep{blandford1979relativistic, Ho2002ApJ}, potentially arising from synchrotron self-absorption near the jet base and radiatively inefficient accretion flows (RIAFs) at low Eddington ratios.

The host galaxy of FRB~20121102A is a low-metallicity, star-forming dwarf with stellar mass $\log(M_*/M_\odot) \approx 8.0$ \citep{Tendulkar2017ApJ}. While uncommon, AGNs in dwarf galaxies have been identified and are thought to host IMBHs \citep{2020ARA&A..58..257G, 2021ApJ...922..155M, 2022NatAs...6...26R,2024MNRAS.527.1962B}. Optical spectra at the PRS location show weak ionization features resembling a LINER-like spectrum \citep{Ho2003ApJ}, consistent with a low-luminosity AGN not strongly influencing its surrounding ISM. The absence of broad emission lines and X-ray non-detections is also compatible with a radiatively inefficient AGN \citep{heckman1980spectral,2023AARv..31....3B}.

From our SED modeling using \texttt{Prospector} (Appendix~\ref{app:prospector}), we estimate a bolometric AGN luminosity of $L_{\rm bol,AGN} \lesssim 1.5 \times 10^{41}$\,erg\,s$^{-1}$. Assuming a central black hole mass $M_{\rm BH} \sim 10^{4.5}\,M_\odot$ \citep{Chen2023ApJ}, this implies an Eddington ratio $\lambda_{\rm Edd} \lesssim 0.04$ using the analytical expression, $\lambda_{\rm Edd} = L_{\rm bol,AGN}/(1.26 \times 10^{38} \left(M_{\rm BH}M_\odot \right))$. This is in line with expectations for radiatively inefficient accretion and is consistent with the properties of known low-excitation radio galaxies (LERGs) \citep{narayan1998advection, hardcastle2007active}.

We also compute the radio-to-X-ray luminosity ratio, defined as $R_X$ = log(LR/LX), using the 5\,GHz flux ($L_R \sim 7 \times 10^{38}$\,erg\,s$^{-1}$) and the 0.5--10\,keV X-ray upper limit ($L_X \lesssim 3 \times 10^{41}$\,erg\,s$^{-1}$; \citealt{Scholz2017ApJ}), yielding $R_X \lesssim -3$. This satisfies the radio-loud criterion defined by \citet{Terashima2003ApJ}, placing PRS1 in the jet-mode AGN regime. Furthermore, the inferred black hole mass and luminosities lie within the radio-loud version of the fundamental plane of black hole activity \citep{Wang2024arXiv240217991W}: $\log L_R = 0.82 \log L_X + 0.07 \log M_{\rm BH} + 5.24$, which predicts $\log L_R \lesssim 39.6$ for $M_{\rm BH} = 10^{4.5} M_\odot$ and $\log L_X \lesssim 41.5$. This is consistent with the observed $L_R \approx 38.8$, suggesting that previous rejections of the AGN hypothesis may have stemmed from applying relations calibrated for radio-quiet sources to a system that is likely radio-loud \citep{Coriat2011, Cao2014}.

An important feature of the AGN scenario is that it allows the FRB engine to be unrelated to the PRS. In this framework, the FRB originates from a separate compact object (e.g., a magnetar) embedded within the nuclear region of the host, while the PRS traces independent AGN activity. This naturally accounts for the disparate variability timescales: PRS1 is stable over years, whereas the FRB’s RM and DM evolve on shorter timescales due to changes in the local plasma environment. Analogous behavior is observed in the Galactic Center magnetar SGR~J1745$-$2900, located $\sim 0.1$\,pc from Sgr~A*, which exhibits high and variable RM consistent with propagation through dense ionized gas in the Galactic nucleus \citep[e.g.,][]{2013Natur.501..391E, 2013ApJ...770L..23M, 2014ApJ...780L...2B}. The co-location of FRB~20121102A and PRS1 to within $\lesssim 40$\,pc \citep{Marcote2017ApJ} permits a similar interpretation.

\begin{figure}[t!]
\centering
\includegraphics[width=0.7\textwidth]{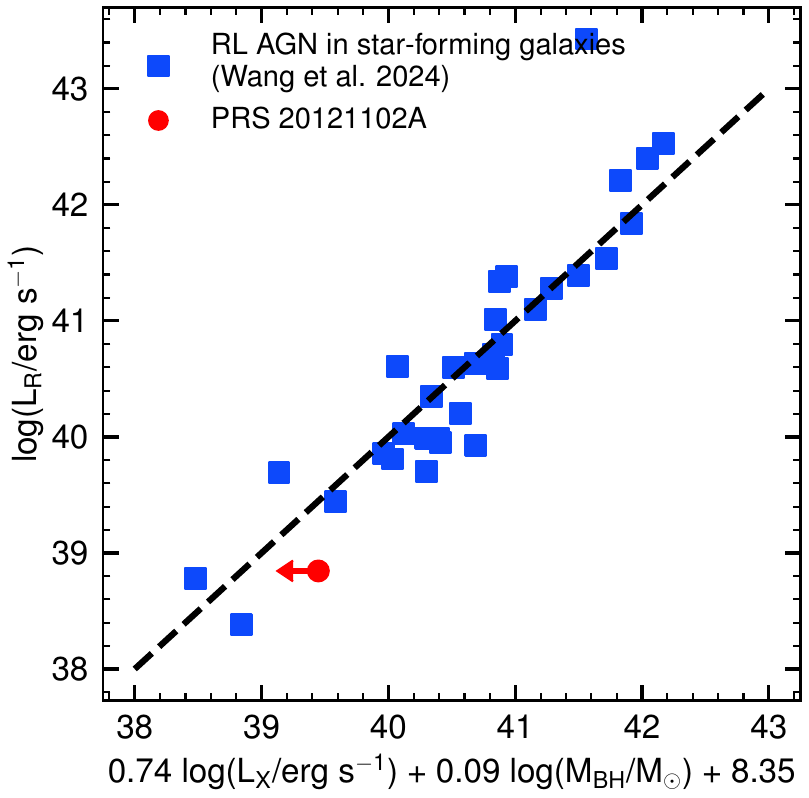}
\caption{PRS1 (red circle) overplotted on the fundamental plane of black hole activity for radio-loud AGNs, adapted from \citet{Wang2024arXiv240217991W}. The source lies within the $1\sigma$ bounds of the expected relation for its estimated X-ray luminosity and black hole mass.}
\label{fig:Fundamental_plane_RLAGN}
\end{figure}

In summary, PRS1 exhibits properties that are broadly consistent with a radio-loud, low-Eddington AGN powered by an intermediate-mass black hole. Its long-term stability, compact morphology, flat radio spectrum, and compatibility with the radio-loud fundamental plane support this interpretation. However, this scenario does not directly account for the FRB itself and requires a physical decoupling between the PRS and the FRB source. While the low-luminosity AGN hypothesis remains viable, current data do not uniquely confirm it, and further high-resolution, multi-wavelength observations will be essential to more definitively evaluate this possibility.

\subsection{The Relationship between Burst Rate and Persistent Radio Source Luminosity}
\label{sec:prs_burst_rate}

Here we investigate whether a correlation exists between the burst rate of active repeating FRBs and their associated PRS luminosity, which offers a crucial test for progenitor models that link burst activity to nebular properties. The motivation for this comparison arises from nebular models of PRSs \citep[e.g.,][]{2017ApJ...839L...3K,2018ApJ...868L...4M, 2020ApJ...896..142B,Sridhar2022ApJ}, which propose that the PRS emission is powered by the same central engine that generates the bursts. In such models, a younger or more active magnetar is expected to simultaneously power both a higher burst rate and a more luminous nebula, potentially imprinting a positive correlation between these observables \citep{2024Natur.632.1014B}.

To test this hypothesis, we select all repeating FRBs with securely identified host galaxies that are either reported on the official CHIME/FRB repeater webpage\footnote{Available at: \url{https://www.chime-frb.ca/repeaters} (last accessed: June 24, 2025).} or have at least one burst detected by CHIME/FRB, such as FRB 20121102A \citep{2019ApJ...882L..18J}. Our sample selection focuses exclusively on repeaters discovered or consistently monitored by CHIME/FRB, specifically utilizing burst rates reported between August 28, 2018, and May 1, 2021 (hereafter referred to as the RN2 window), as presented in the second CHIME/FRB repeater catalog \citep{2023ApJ...947...83C}. This restriction minimizes selection biases arising from varying telescope sensitivities, observing frequencies, and follow-up strategies, thereby ensuring a more uniform treatment of burst detectability across the sample. Notably, this excludes FRB 20190520B \citep{2022Natur.606..873N}, as CHIME/FRB had no exposure at its coordinates during the RN2 window.

For the three new active repeaters—FRBs 20220912A, 20240114A, and 20240209A \citep{2024MNRAS.529.1814H,2025arXiv250611915B,2025ApJ...979L..21S}—that were discovered after the RN2 window, we calculate 1$\sigma$ upper limits on their burst rates using the formalism employed by \cite{2023ApJ...947...83C}. To do so, we utilize the publicly available RN2 exposure map \citep{2023ApJ...947...83C} and their respective 95\% completeness fluence limits: 15.65 Jy\,ms for FRB 20240114A and 1.5 Jy\,ms for FRB 20240209A \citep{2025arXiv250513297S,2025ApJ...979L..21S}. For FRB 20121102A, we adopt a 95\% completeness fluence limit of 7 Jy\,ms from \cite{2019ApJ...882L..18J}. For FRBs 20220912A and 20200120E, for which specific fluence completeness limits are not available, we adopt the Catalog-1 mean fluence completeness limit of 5 Jy\,ms \citep{catalog12021ApJS}.

We then compile the spectral radio luminosities of the associated PRSs at a common frequency of 3 GHz, as detailed in Table \ref{tab:frb_prs_properties}. When 3 GHz flux densities are not directly available in the literature, we interpolate the published values using reported spectral indices. For sources without a reported spectral index, we adopt $\alpha_{\nu} = -0.4$ (i.e., $L_\nu \propto \nu^{-0.4}$), a choice motivated by the observed spectral indices of PRSs associated with FRBs 20190520B and 20240114A. We note that our results are not sensitive to the precise choice of spectral index; for instance, adopting a flat spectrum ($\alpha_{\nu} = 0$), consistent with FRB 20121102A, yields similar results.

As discussed by \cite{2024ApJ...971L..51B} and \cite{2025arXiv250601238L}, to correct for cosmological effects, we convert the fluence thresholds for each source into minimum detectable isotropic energies using their host redshifts and the relation $E_{\mathrm{min}} = 4\pi D_L^2(z) F_\nu \delta\nu$, assuming $\delta\nu = 400$ MHz, consistent with values adopted in the CHIME/FRB Catalog 1 \citep{catalog12021ApJS}. All burst rates are then rescaled to a common fiducial energy threshold of $10^{39}$ erg using $R(E) \propto E^\gamma$ with $\gamma = -1.5$, as employed by \cite{2023ApJ...947...83C}. For FRB 20121102A, we adopt a redshift of $z=0.19273$ from \cite{Tendulkar2017ApJ}. The compiled list of FRBs, along with their redshifts, PRS luminosities, burst rates, and relevant references are presented in Table~\ref{tab:frb_prs_properties}.

\begin{figure}[ht]
    \centering
    \includegraphics[width=0.75\textwidth]{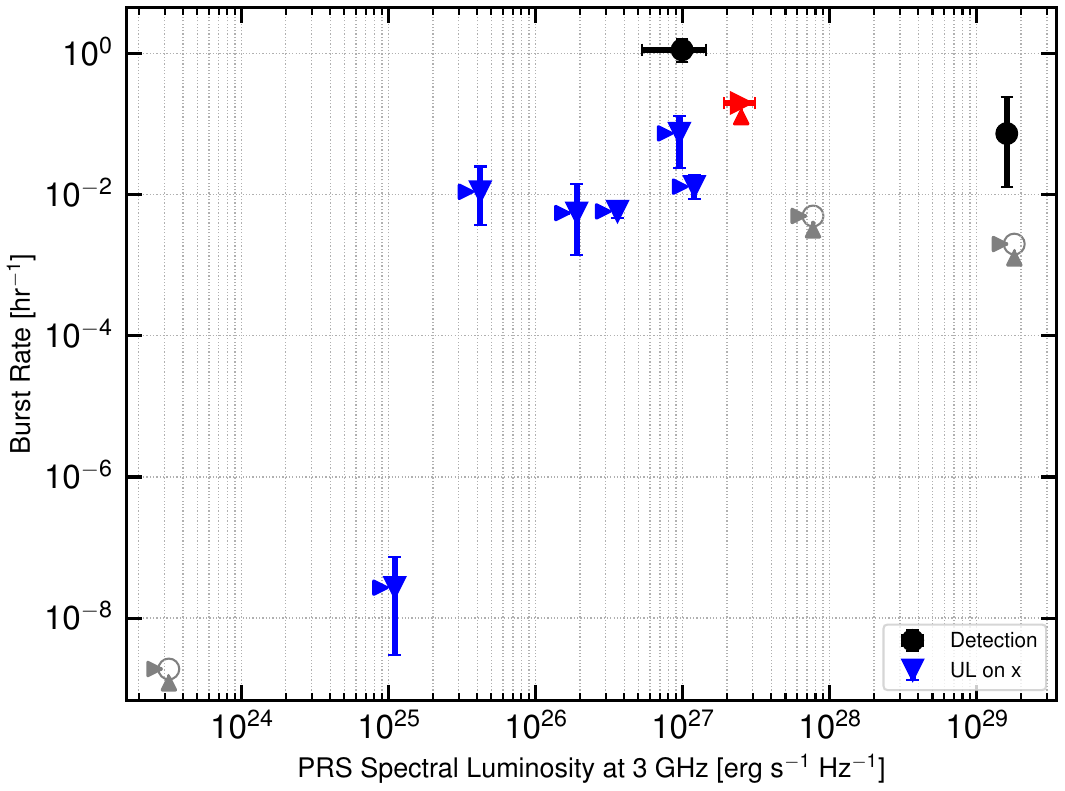}
    \caption{
        Scatter plot showing the burst rate (in hr$^{-1}$) as a function of the spectral radio luminosity at 3 GHz (in erg s$^{-1}$ Hz$^{-1}$) for repeating FRBs with known host galaxies. Black circles represent sources where both the burst rate and PRS luminosity are measured with 1$\sigma$ uncertainties. Blue triangles denote sources with upper limits on the PRS luminosity but measured burst rates. Red triangles show the opposite case: upper limits on the burst rate with known luminosities. Gray open circles with arrows represent upper limits on both quantities. Arrows indicate the direction of the limits. See \S\ref{sec:prs_burst_rate} for discussion.
    }
    \label{fig:prs_burst_rate}
\end{figure}

In Figure~\ref{fig:prs_burst_rate}, we plot the burst rate versus the 3 GHz spectral luminosity of the associated PRS. The plot includes a mixture of measured values and upper limits, indicated by error bars and arrows, respectively. A visual inspection reveals no clear correlation. To assess this statistically, we apply the censored Kendall’s $\tau$ test \citep{1996MNRAS.278..919A}, using the publicly available Python implementation developed by \citet{2022ApJ...930..126F}, which properly accounts for upper limits in both variables. The resulting coefficient is $\tau \approx 0.18$ with a p-value of 0.41, indicating no statistically significant correlation between PRS luminosity and burst activity.

Using the reported CHIME/FRB burst rates for FRBs 20220912A, 20240114A, and 20240209A from the post-RN2 period \citep{2022ATel15679....1M,2025arXiv250513297S,2025arXiv250611915B,2025ApJ...979L..21S} would yield an even more scattered distribution, further supporting the absence of a clear correlation. This finding is in tension with scenarios where all active repeaters are very young magnetars \citep[e.g., $\lesssim$ 100 years old;][]{2025arXiv250601238L} whose burst activity and nebula luminosity are tightly coupled, as posited by some young neutron star progenitor models.

Finally, we caution that our analysis is subject to important limitations stemming primarily from the complex and diverse nature of FRB repetition. Burst rates are known to be highly time-variable and often deviate significantly from a simple Poisson distribution, exhibiting clustering and quiescent periods \citep[e.g.,][]{2016MNRAS.458L..89C,2018MNRAS.475.5109O,2022ApJ...927...59L}. This inherent non-Poissonian behavior, coupled with varying activity patterns across different repeaters (where some show consistent Poissonian rates, while others display extreme variability), makes precise determination of a ``true" average burst rate challenging. Furthermore, our reliance on specific observing windows (RN2) and calculated upper limits for some sources, while necessary to minimize selection biases, means the sample is still limited by current observational capabilities and finite monitoring durations. These factors can potentially obscure underlying physical correlations. Future efforts with long-term, homogeneous monitoring campaigns and deeper radio imaging of repeaters across a wider range of redshifts and host galaxy environments will be crucial. Such data will enable more robust characterization of burst activity, better constraints on PRS energetics, and ultimately, a more definitive assessment of any connection between burst rate and PRS luminosity, offering key insights into the elusive central engine of FRBs.

\begin{table*}[ht!]
\begin{center}
\caption{Properties of Our Sample of Repeating FRB Sources with Host Galaxy Associations}
\label{tab:frb_prs_properties}
\resizebox{\textwidth}{!}{%
\begin{tabular}{ccccc}
\hline
FRB Source & Redshift ($z$) & $L_{\text{3GHz}}$ (erg s$^{-1}$ Hz$^{-1}$)$^{a}$ & Burst Rate (hr$^{-1}$) & Key References \\
\hline
20121102A & 0.1927 & $(1.6\pm 0.2) \times 10^{29}$ & $0.07^{+0.17}_{-0.06}$ & \cite{Tendulkar2017ApJ,Marcote2017ApJ,2024ApJ...971L..51B} \\
20180814A & 0.0684 & $< 9.5 \times 10^{26}$ & $0.07_{-0.05}^{+0.06}$ & \cite{2023ApJ...950..134M,2023ApJ...947...83C} \\
20180916B & 0.0337 & $< 3.6 \times 10^{26}$ & $0.0060_{-0.0011}^{+0.0013}$ & \cite{Marcote_2020_Natur,2023ApJ...947...83C} \\
20181030A & 0.0037 & $< 1.1 \times 10^{25}$ & $(2.7_{-2.4}^{+4.6}) \times 10^{-8}$ & \cite{2021ApJ...919L..24B,2023ApJ...947...83C} \\
20190110C & 0.1224 & $< 1.9 \times 10^{26}$ & $0.006_{-0.004}^{+0.009}$ & \cite{2024ApJ...961...99I,2023ApJ...947...83C} \\
20190303A & 0.0634 & $< 1.2 \times 10^{27}$ & $0.013_{-0.004}^{+0.006}$ & \cite{2023ApJ...950..134M,2023ApJ...947...83C} \\
20190520B & 0.241 & $(2.4\pm 0.4) \times 10^{29}$ & $--$ & \cite{2022Natur.606..873N} \\
20200120E & 0.00082 & $< 3.2 \times 10^{23}$ & $<1.9 \times 10^{-9} $ & \cite{2021ApJ...910L..18B,Kirsten_2022_Natur}; this work \\
20200223B & 0.06024 & $< 4.2 \times 10^{25}$ & $0.011_{-0.007}^{+0.014}$ & \cite{2024ApJ...961...99I,2023ApJ...947...83C}\\
20201124A & 0.0980 & $(9.9\pm 4.6) \times 10^{26}$ & $1.1_{-0.4}^{+0.5}$ & \cite{2022ApJ...927...59L,2024Natur.632.1014B} \\
20220912A & 0.0771 & $< 7.7 \times 10^{27}$ & $<0.005$ & \cite{2023ApJ...949L...3R,2024MNRAS.529.1814H} \\
20240114A & 0.1300 & $(2.5\pm 0.6) \times 10^{27}$ & $<0.2$ & \cite{2025arXiv250611915B,2025AA...695L..12B} \\
20240209A & 0.1384 & $< 1.8 \times 10^{29}$ & $<0.002$ & \cite{2025ApJ...979L..21S}; this work \\
\hline
\end{tabular}
}

$^a$ Reported PRS luminosities at 3 GHz are either directly measured or extrapolated from lower-frequency data using published spectral indices. When unavailable, we adopt $\alpha_{\nu} = -0.4$ as a conservative assumption (see \S \ref{sec:prs_burst_rate}).\\
\end{center}
\end{table*}

\subsection{Future Directions and Open Questions}
\label{subsec:future_directions}

Our results provide the most stringent constraints to date on the long-term variability of PRS1, showing stable L-band flux over a 7-year baseline with variability fully consistent with interstellar scintillation. This places important limits on models where the persistent emission is intrinsically coupled to a rapidly evolving outflow. To further distinguish between competing scenarios—such as a magnetar wind nebula, hypernebula, or a low-luminosity active galactic nucleus—several key observational efforts are needed.

Continued long-term monitoring across a broad range of frequencies remains essential. While the L-band provides the richest historical dataset and thus served as the focus of this work, observations at both lower (e.g., Band 3) and higher (e.g., Ku- and K-band) frequencies will help constrain spectral evolution, variability amplitude, and possible low-frequency spectral turnovers. Our new 745\,MHz detections show no evidence for significant modulation beyond that expected from RISS, and the inverse-variance weighted mean flux is consistent with a flat or mildly declining spectrum. The 400\,MHz non-detections—though not deep enough to constrain the known low-frequency fluxes from earlier GMRT epochs \citep{2020MNRAS.498.3863M,2021AA...655A.102R}—highlight the need for deeper measurements below 600\,MHz. Instruments like LOFAR, GMRT, and MWA could test for low-frequency curvature or absorption mechanisms, such as free-free absorption by dense plasma near the source.

High-resolution VLBI imaging will also be critical for tracking the spatial morphology of PRS1. Any detectable expansion, change in size, or asymmetry could directly constrain the age and physical mechanism powering the emission \citep[e.g.,][]{2022MNRAS.511.6033P}. Multi-epoch VLBI campaigns will complement variability studies and provide independent diagnostics of the source geometry.

Improved X-ray limits, particularly with high-sensitivity instruments like Chandra or XMM-Newton, would tighten constraints on radiative efficiency and the Eddington ratio—key diagnostics for assessing an LLAGN origin. Moreover, deep optical or near-infrared spectroscopy of the host could reveal weak AGN indicators such as coronal lines or broadened features that might remain undetected in lower-resolution data. Such data would also help distinguish between central and off-nuclear FRB origins and better characterize the local environment.

As the population of FRBs with persistent counterparts grows, systematic comparisons of their radio light curves, spectral indices, and host galaxy demographics will become increasingly powerful. Notably, a subset of repeating FRBs, including FRB~20121102A, are now known to reside in star-forming dwarf galaxies—precisely the environments where IMBHs are predicted to reside \citep{2022NatAs...6...26R}. This opens the possibility that some PRSs may trace a population of previously undetected IMBHs, providing a new observational pathway for probing black hole seed formation and post-merger dynamics \citep{2020ApJ...888...36R,2020ApJ...895...98E}.

An intriguing clue comes from the observed $\sim$159-day periodic activity cycle of FRB~20121102A \citep{2020MNRAS.495.3551R,2021MNRAS.500..448C,2025A&A...693A..40B}. This timescale—possibly related to orbital motion, jet precession, or circumburst environmental modulation—has yet to be incorporated into most PRS models. Its apparent disconnect from the stable radio emission further motivates scenarios where the FRB engine is physically distinct from the PRS. Indeed, the combination of stable persistent flux, significant RM and DM variability, and ultra-short FRB burst durations are difficult to reconcile within a unified, nebula-powered framework. Instead, they may point toward a scenario in which PRS1 is an LLAGN powered by an IMBH, while the FRB activity originates from a separate compact object in the same galactic environment. This naturally explains the observed temporal decoupling between burst activity and persistent emission (see \S\ref{subsubsec:agn_scenario}).

Ultimately, understanding PRSs—whether as magnetized nebulae, relativistic shocks, or dormant AGNs—offers a unique window into the diverse environments and evolutionary paths of FRB sources. As new telescopes bring greater sensitivity, resolution, and frequency coverage, targeted long-term monitoring of nearby systems like FRB~20121102A will remain pivotal for advancing both FRB and compact object astrophysics.

\section{Conclusion}
\label{sec:conclusion}

In this paper, we presented a comprehensive study of the long-term radio evolution of the persistent radio source associated with FRB~20121102A, leveraging new uGMRT observations alongside extensive archival data spanning seven years (2016--2023). Our primary observational finding is the remarkable long-term stability of PRS1's L-band flux density, with no statistically significant trend detected over this period. The inverse-variance weighted mean flux density is $213 \pm 4\,\mu$Jy. While we detect statistically significant variability at 1.4\,GHz, Monte Carlo simulations show that this is fully consistent with the RISS combined with measurement noise, indicating that PRS1 is not intrinsically variable at these timescales. We extended our analysis to lower frequencies using new uGMRT Band 4 measurements and archival data at 610--745\,MHz. We find no evidence for significant variability in this band, with a modulation index ($0.11 \pm 0.08$) consistent with RISS expectations. The inverse-variance weighted mean flux density at 745\,MHz is $246 \pm 16\,\mu$Jy, yielding a spectral index of $\alpha = -0.15 \pm 0.08$ when compared with the L-band mean, confirming a flat radio spectrum. New Band 3 observations at 400\,MHz result in non-detections with $3\sigma$ upper limits of $\sim275$--$295\,\mu$Jy. These are consistent with earlier detections but do not improve existing constraints.

These observational characteristics impose strong constraints on proposed models for PRS1. The absence of a long-term decline disfavors models involving rapidly evolving outflows, such as long GRB afterglows, superluminous supernovae, or expanding supernova ejecta. Similarly, the predicted secular evolution from both MWN and hypernebula models may be in mild tension with the observed flatness of the PRS1 light curve. In the MWN case, resolving this tension may require the source to be significantly older than inferred from RM evolution. For hypernebula models, the predicted flux increase is not seen in our data, though modest adjustments to the input parameters could reconcile the model with observations. Finally, we argued that the AGN scenario remains a viable interpretation for PRS1. The observed stability, compact morphology, flat spectrum, and consistency with the radio-loud AGN fundamental plane (assuming an intermediate-mass black hole) are all consistent with a low-Eddington, jet-mode AGN. This interpretation naturally decouples the persistent and burst emission, helping to reconcile their contrasting variability timescales. However, the evidence remains circumstantial, and distinguishing between AGN and other scenarios will require continued high-resolution, multi-wavelength monitoring.

The detection of ultra-fast bursts with microsecond substructure from FRB~20121102A favors models in which FRB emission originates in extremely compact regions near the neutron star magnetosphere, disfavoring far-field emission models invoked in nebular or shock-powered scenarios. Additionally, using updated luminosities and homogeneous burst rate estimates from the CHIME/FRB RN2 observing window, we find no statistically significant correlation between the burst rate and PRS luminosity. This disfavors simple one-to-one models in which a common energy reservoir powers both the bursts and the nebular emission, and instead supports scenarios where the FRB and PRS may arise from physically distinct regions or decoupled energy reservoirs.

Although PRS1 may not directly reveal the FRB progenitor, its stability, spectrum, and compactness offer valuable insight into the environments of repeating FRBs. Continued long-term and multi-frequency observations, coupled with population studies of other persistent counterparts, will be key to constraining the diversity of FRB origins.

\begin{acknowledgments}

We thank the staff of the GMRT that made these observations possible. GMRT is run by the National Centre for Radio Astrophysics of the Tata Institute of Fundamental Research. The National Radio Astronomy Observatory is a facility of the National Science Foundation operated under cooperative agreement by Associated Universities, Inc. We acknowledge use of the CHIME/FRB Public Database, provided at \url{https://www.chime-frb.ca/} by the CHIME/FRB Collaboration. M.B is a McWilliams fellow and an International Astronomical Union Gruber fellow. M.B. also receives support from the McWilliams seed grant. 
\end{acknowledgments}

\vspace{5mm}
\facilities{GMRT, VLA}

\software{astropy \citep{2013AA...558A..33A},  
          dynesty \citep{2020MNRAS.493.3132S}, 
          matplotlib \citep{2007CSE.....9...90H},
          scipy \citep{2020NatMe..17..261V},
          casa \citep{2007ASPC..376..127M},
          kendall \citep{Flury_kendall},
          science-plots \citep{SciencePlots},
          numpy \citep{2020Natur.585..357H},
          FSPS \citep{Conroy2009ApJ},
          Prospector \citep{Leja2017ApJ,prospect2019},          
          }

\appendix

\section{Stellar Population Synthesis Using {\tt Prospector}}
\label{app:prospector}

To determine key properties of the host galaxy of FRB~20121102A, including its stellar population characteristics and the bolometric luminosity of the putative AGN, we utilize \texttt{Prospector} \citep{johnson:17,Johnson:21}. This Python-based Bayesian inference code generates model spectral energy distributions (SEDs) using stellar population synthesis models defined within the Flexible Stellar Population Synthesis (FSPS) framework \citep{Conroy2009ApJ}, incorporating the MILES stellar library and MIST isochrones.

We perform a comprehensive spectrophotometric fitting of 8 archival photometric data points from the Pan-STARRS, HST and Spitzer data (see Table \ref{tab:sed_photometry}) and spectroscopic data from the Gemini Multi-Object Spectrograph (GMOS) of the host galaxy which was reduced by \cite{Tendulkar2017ApJ}. Our fitting employs a flexible eight-bin non-parametric star formation history (SFH) model \citep{suess:21,suess:22}, a \citet{chabrier:09} initial mass function (IMF), and the \citet{Kriek_2013ApJ...775L..16K} dust law. Posterior sampling is conducted using the \texttt{dynesty} nested sampling algorithm \citep{2020MNRAS.493.3132S}. We fix the redshift to $z = 0.19273$ and apply \citet{gallazzi:05} stellar mass-stellar metallicity relationship in the local universe as a Gaussian prior. To improve constraints on star-formation rates, we also adopt UniverseMachine priors \citep{2019MNRAS.488.3143B} on SFR ratios in each bin, which generally yield more robust results than constant SFR assumptions.

Crucially for our AGN analysis, we allow for the presence of an AGN component, primarily contributing to the mid-infrared emission, using the CLUMPY AGN templates described in \cite{prospect2019}. This model incorporates two key parameters: $f_{agn}$ (the ratio of AGN luminosity to bolometric stellar luminosity) and $\tau_{agn}$ (the optical depth of the AGN torus dust). In this model, the majority of UV/Optical emission from the central engine is assumed to be obscured by the AGN dust torus, with any leak-out further attenuated by the galaxy dust attenuation model. We adopt a log-uniform prior for both $f_{agn}$ ($10^{-5} < f_{agn} < 3$) and $\tau_{agn}$ ($5 < \tau_{agn} < 150$). These broad ranges are chosen to ensure data-driven posteriors rather than prior-driven results, and they are consistent with the observed power-law distribution of black hole accretion rates \citep{2018MNRAS.480.4379C}.

In addition to the AGN parameters, our fitting simultaneously estimates host galaxy stellar population properties, including stellar mass, star-formation rate (SFR), mass-weighted ages ($Age_{MW}$), stellar metallicity, and dust content ($A_V$). Table \ref{tab:appendix1} presents the recovered 16th, 50th, and 84th percentile posterior distributions for all these properties.

\begin{deluxetable}{cccccc}[]
\tablecaption{AGN parameters and host galaxy posterior distributions obtained from Prospector SED modeling${}^*$
\label{tab:appendix1}}
\tablehead{\colhead{No} & \colhead{Property} & \colhead{Units} & \colhead{16th} & \colhead{50th} & \colhead{84th} \\ {} & {} & {} & {} & {} & {}}
\startdata
1 & $f_{agn}$ & $L_{AGN}/L_*$ & 0.0011 & 0.0087 & 0.0288 \\
2 & $\tau_{agn}$ & optical depth & 13.17 & 86.92 & 121.15 \\
3 & $L_{bol}$ & $\log(L_{\odot})$ & 9.14 & 9.36 & 9.51 \\
4 & $L_{bol, AGN}$ & $\log(L_{\odot})$ & 6.34 & 7.01 & 7.27 \\
5 & Stellar Mass & $\log(\mathrm{M_*/M_{\odot}})$ & 7.8 & 8.0 & 8.2 \\
6 & SFR & M$_{\odot}$/yr & 0.11 & 0.21 & 0.29 \\
7 & $Age_{MW}$ & Gyr & 0.9 & 1.2 & 1.5 \\
8 & Stellar Metallicity & $\log(\mathrm{Z_*/Z_{\odot}})$ & -1.13 & -0.68 & -0.42 \\
9 & Av & mag & 0.2 & 0.4 & 0.6 \\
\enddata
\end{deluxetable}

\begin{figure*}
\centering
\includegraphics[width=0.95\textwidth]{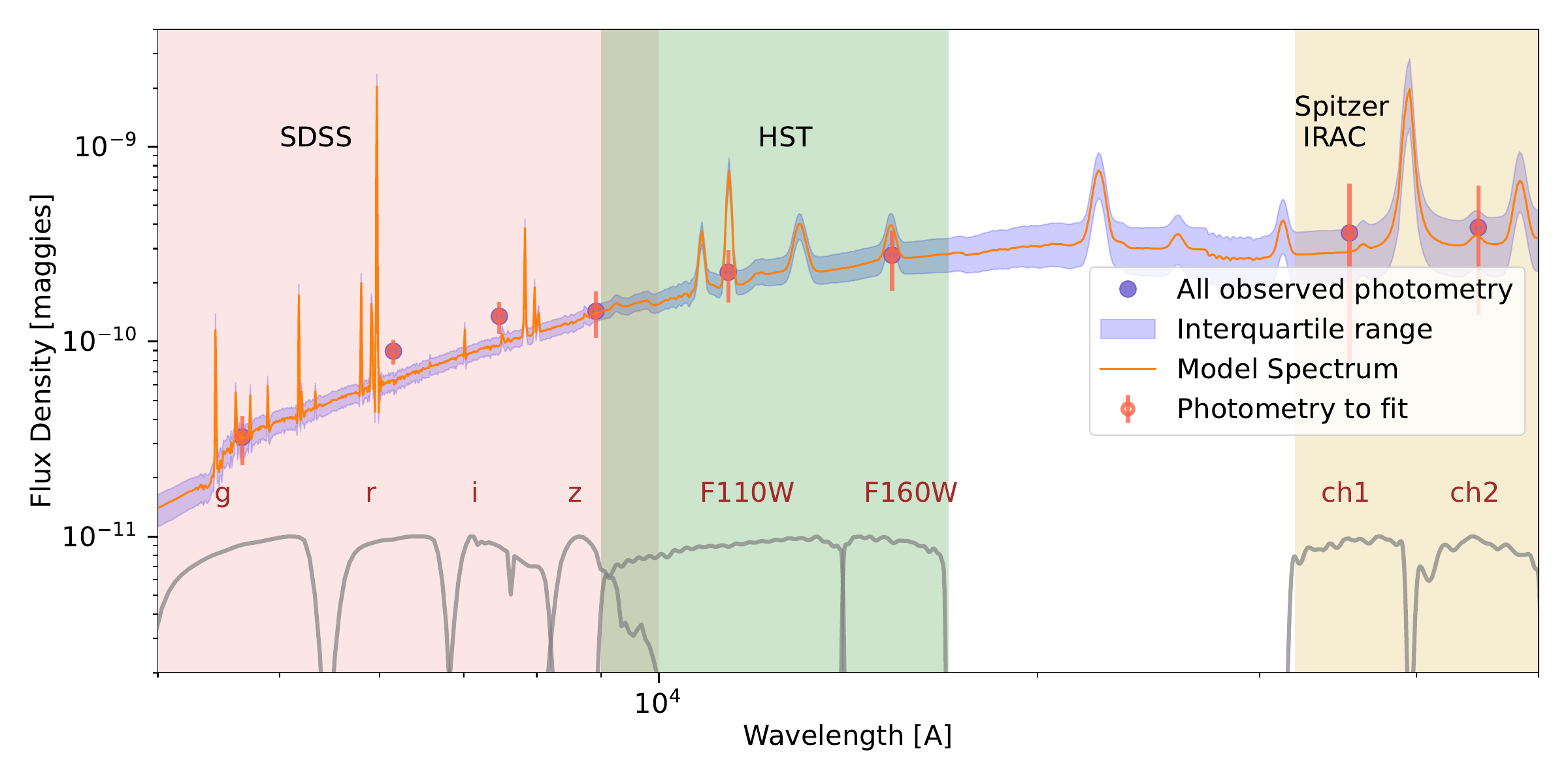}
\caption{Modelling the SED of the FRB 20121102A host. The flux density of the FRB 20121102A host, a low-metallicity ($\sim$ 0.2Z$_{\odot}$) star-forming ($\sim$ 0.2 M$_{\odot}$/yr) dwarf galaxy ($\sim 10^{8}$ M$_{\odot}$), in optical and near-IR bands are plotted along with the best-fit {\tt Prospector} model spectrum. The actual photometry data are also shown to assess the quality of the {\tt Prospector} model.}
\label{fig:SEDS_1}
\end{figure*}

\begin{table}[ht]
\begin{center}
\caption{Broadband photometry used for SED fitting.}
\label{tab:sed_photometry}
\begin{tabular}{@{}llll@{}}
\toprule
Instrument & Filter & Effective Wavelength [\AA] & Flux Density$^{a,b}$ [maggies] \\
\midrule
SDSS & g & 4686 & 4.57 $\times 10^{-11}$ \\
     & r & 6165 & 6.55 $\times 10^{-11}$ \\
     & i & 7481 & 1.26 $\times 10^{-10}$ \\
     & z & 8931 & 1.91 $\times 10^{-10}$ \\
HST/WFC3 & F110W & 11534 & 3.39 $\times 10^{-10}$ \\
         & F160W & 15369 & 4.74 $\times 10^{-10}$ \\
Spitzer/IRAC & Ch1 & 35634 & 2.84 $\times 10^{-10}$ \\
             & Ch2 & 45110 & 2.49 $\times 10^{-10}$ \\
\bottomrule
\end{tabular}
\end{center}
\vspace{-1ex}
\begin{flushleft}
$^{a}$ 1 maggie is defined as the flux density in Jy divided by 3631. Fluxes are corrected for Galactic extinction following \citet{2011ApJ...737..103S}. The reported fluxes are taken from \cite{2017ApJ...843L...8B}. \\
$^{b}$ All broadband fluxes are assigned a 20\% fractional uncertainty. \\
\end{flushleft}
\end{table}

\bibliography{sample631}{}
\bibliographystyle{aasjournal}

\end{document}